# DISPERSION IMPACT ON BALLISTIC CNTFET N+-I-N+ PERFORMANCES

*Patricia Desgreys, Jefferson Gomes Da Silva, Dorothée Robert*

LTCI – CNRS 5818 – GET-Télécom Paris
46 rue Barrault
75634 Paris Cedex 13 - France

**ABSTRACT**

The designs integrating the promising carbon nanotube transistors (CNTFET) will have to take into account the constraints implied by the strong dispersion inherent to nanotube manufacturing. This paper proposes to characterize the main CNTFET performances: on-current $I_{on}$, $I_{on}/I_{off}$ ratio and inverse sub-threshold slope S according to the dispersion on the nanotube diameter. For this purpose, we use a compact model suitable for testing several diameter values.

## 1. INTRODUCTION

The carbon nanotubes belong to the fullerenes family which is a carbon allotropic variety. Since their discovery in 1985, their remarkable structure and their properties have been studied and recently, carbon nanotube field effect transistors have attracted a great deal of attention as possible building blocks of a future nanoelectronics due to their unique structural and electronic properties.

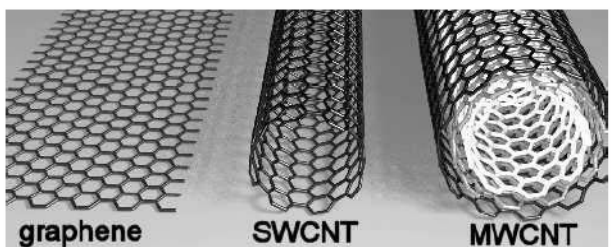

**Figure 1. Carbon nanotubes.**

A carbon nanotube can be seen like a graphite sheet rolled up on itself with a diameter typically between 1 and 5 nm. The nanotube can be a metal or a semiconductor according to its chirality, characterized by the chirality vector (n, m). Because only semiconductor nanotubes are appropriate for transistors implementation, we initially consider semiconductors single walled nanotubes obtained by free growth. The assumed chirality is often the zigzag one (n, 0) which ensures the semiconductor character provided that n is not a multiple of three. It results from this geometry a strong 1D quantum confinement around the circular direction of the tube which takes part in the good electronic properties of the nanotubes: high carrier mobility, low leakage current, important on state current relatively to the applied voltages and low inverse sub-threshold slope. These properties allow us to consider the design of high-speed and high performances electronic circuits.

The manufacturing processes progress continually: one of the most advanced is the Chemical Vapor Deposition technique (CVD). However, there is still a strong spreading on the obtained diameters and the control or the selection of the diameters and/or the chiralities remain complex operations. Finally small semiconductors nanotubes are easier to manufacture with accuracy.

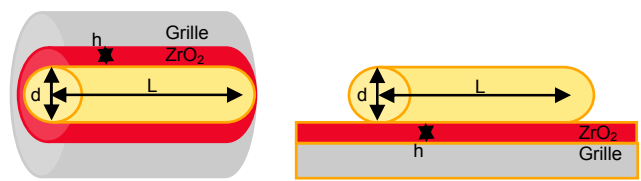

**Figure 2. CNTFET geometry and key parameters.**

The central part of a CNTFET is an intrinsic (i) semiconductor tube surrounded with or posed on a combination insulator plus gate; this assembly does not have specific n or p behavior but presents a total ambivalent behavior according to the voltage applied to the gate. The three key parameters which govern the CNTFET operation are:
- the length: we consider low length tubes (L<15nm) so that transport in the nanotube is ballistic that corresponds to the statistical absence of collisions and thus to the maximum carrier speed.
- the diameter : the current delivered by the transistor according to the voltages which are applied to it strongly depends on the nanotube's diameter which fixes the energy of the bandgap $E_g$ and the number of the implied sub-bands.
- the nature of the contacts at each extremity of the intrinsic tube: the current is controlled by the

transmission of the carriers through the contacts. The two types mostly considered, are schottky contacts metal-i-metal and doped nanotubes n+-i-n+; in principle, the second one exhibits superior device characteristics (better off state and better on state) [1],[2].

Now, the improvement of the CNTFET must be coupled with the development of circuit compatible models so that the possibilities of these new components can be explored at the system level. In this context, we present our work on compact modeling and simulation of the ballistic CNTFET for several diameters. This model will allow to test electronic architectures taking into account component dispersion.

Part 2 details the compact model used and our contribution to the widening of this model to any value of d ranging between 1 and 5 nm. Part 3 shows simulation results obtained from a MATLAB implementation of this model and part 4 studies the impact of diameter, carbon-carbon bonding energy and oxide dielectric constant dispersions on the transistor static performances.

## 2. MODEL PRESENTATION

### 2.1. 1D ballistic model

Ballistic doped drain/source CNTFET operates like a conventional MOSFET with a barrier height modulation and there is no tunnelling barrier. Its physical structure is an intrinsic CNT with n+ doped nanotubes as a source and drain. A coaxial gate is placed around the intrinsic part of the nanotube and separated by an oxide. The model exposed here is in a large part extracted from the paper [3]. It is based on the works of the most advanced research group with Datta and Lundstrom from Purdue University and this model is close to underlying physics.

The gate voltage $V_G$ induces charge $Q_{CNT}$ and potential $\Psi_S$ in the CNTFET channel. It also modulates the top barrier of the energy band between the source and the drain. As the source-drain barrier is lowered, current flows between the source and the drain. For ballistic transport, all scattering mechanisms can be neglected. The current will remain constant throughout the channel and we may compute it at the top of the energy band, at the channel beginning.

### 2.2. Model equations

The relation between $\Psi_S$ and $V_G$ is given by:

$$\psi_S = V_G - \frac{Q_{CNT}}{C_{INS}} = V_G - \frac{e}{C_{INS}} \cdot \sum_p \left[ n_{Sp}(\psi_S) + n_{Dp}(\psi_S) \right] \quad (1)$$

$n_{Sp}$ and $n_{Dp}$ are the carrier's densities in the sub-band p and coming respectively from the source and the drain.

The calculation of these quantities requires the knowledge of the potential $\Psi_S$.

$$n_{ip} = N_0 F(\Delta_p, \xi_i) \quad i = S, D \quad \xi_i = \frac{\psi_S - \Delta_p - \mu_i}{K_b T} \quad (2)$$

$\Delta_p$ is the energy of the bottom of the sub-band p conduction band.

$$F(x,y) = \int_0^{+\infty} \frac{N_0}{1 + \exp(\frac{\sqrt{z^2 + y^2} - y}{K_b \cdot T} - x)} dz \quad (3)$$

$$z = \sqrt{(E + \Delta_p)^2 - \Delta_p^2} \qquad N_0 = \frac{4 K_b T}{3 \pi V_\Pi b} \quad (4)$$

$V_\Pi$ is the carbon-carbon bonding energy and b the C-C bonding distance. Then, the knowledge of the potential $\Psi_S$ allows the calculation of the drain source current according to the Landauer formula:

$$I_{DS} = \frac{4 e K_b T}{h} \sum_p \left[ \ln(1 + \exp(\xi_s)) - \ln(1 + \exp(\xi_D)) \right] \quad (5)$$

The difficulty is that the equations (1) and (2) must be self consistently solved. However F(x, y) does not have an analytical solution. A possibility is then to approach this equation by the two following functions (valid for $\Delta p < 0.5$ eV):

$$\frac{n}{N_0} = A \exp(\xi), \quad \text{for } \xi < 0$$
$$= B\xi + C, \quad \text{for } \xi \geq 0 \quad (6)$$

with $A = C = -5.3\Delta_p^2 + 10\Delta_p + 1$ and $B = 0.34\Delta_p + 1$

With this approach, we can numerically solve the model equations set and find a relation between $V_G$ and $\Psi_S$. In this resolution, the nanotube's diameter d is implied in the calculation of the bottom energy of the first conduction sub-band $\Delta_1$ (7), in the determination of the number of sub-bands necessary for the current calculation (8) and in the expression of the insulator capacity $C_{INS}$ (9) or (10).

$$\Delta_1 = \frac{V_\Pi \cdot b}{d} = \frac{0.43}{d[nm]} \quad (= E_g/2) \quad (7)$$

$$\Delta_p = \Delta_1 \frac{6p - 3 - (-1)^p}{4} \quad \text{(in the vicinity of } E_F\text{)} \quad (8)$$

If the nanotube is posed on an oxide plane of height h, of permittivity $\varepsilon_{ox}$, the capacity per length unit is expressed by [4]:

$$C_{INS} = \frac{2\pi \varepsilon_0 \varepsilon_{ox}}{\ln\left[2(2h+d)/d\right]} \quad (9)$$

If the intrinsic nanotube is surrounded by a coaxial gate, both separated by an oxide of thickness h, the gate influence is more important (traditional cylindrical structure):

$$C_{INS} = \frac{2\pi \varepsilon_0 \varepsilon_{ox}}{\ln\left[(2h+d)/d\right]} \quad (10)$$

The equations were solved in a particular case of diameter and value of insulator capacitance. The results extracted from [3] and reported in figure 3, show that there is an empirical approximate relation between potential $V_G$ and $\Psi_S$:

$$V_{GS} - \psi_S = 0, \quad \text{pour } V_{GS} < \Delta_1 \quad (11)$$
$$= \alpha(V_{GS} - \Delta_1), \quad \text{pour } V_{GS} \geq \Delta_1$$

$$\alpha = \alpha_0 + \alpha_1 VDS + \alpha_2 VDS^2 \quad (12)$$

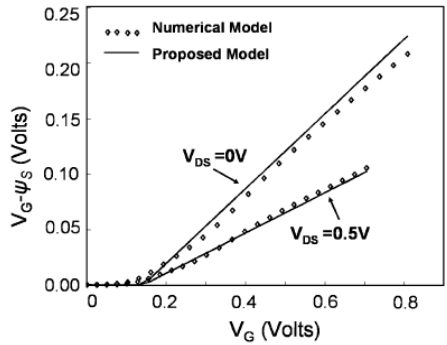

**Figure 3.** $V_G$-$\Psi_S$ in function of $V_G$ for two values of $V_{DS}$ for d=3nm and $C_{INS}$=17 pF/cm [3].

This approximation is interesting because it allows to quickly calculate the current in the transistor starting from the voltages applied to the gate and the drain (by taking the source as reference) and the data of the three empirical parameters $\alpha_0$, $\alpha_1$ and $\alpha_2$.

### 2.3. Automatic determination of parameters $\alpha_0$, $\alpha_1$ and $\alpha_2$

From there, our work focused on the development of a MATLAB program implementing the automatic calculation of the values $\alpha_0$, $\alpha_1$ and $\alpha_2$. The program CMVD (Compact Model for Variable Diameter) inputs are the transistor geometry, the oxide nature and thickness and it delivers $\alpha_0$, $\alpha_1$ and $\alpha_2$ in function of d in the range from 1 to 5 nm. The MATLAB program solves the equations for a vector of $V_{GS}$ values and for three values of $V_{DS}$. Initially, a great number of $V_{GS}$ were considered and $\alpha$ was the slope of least squares straight line. This phase of analysis allow to check that the linear behaviour was well observed as on figure 3 for other values of d and $C_{INS}$. So we simplified the program considering only two values of $V_{GS}$ for the determination of $\alpha$ : $V_{GS}$ =0 and $V_{GS}$ =0.8V. The set of values obtained can be introduced into a compact model of ballistic transistor CNTFET n+-i-n+ implementing the equations (2), (5), (7), (8), (11) and (12). We can thus quickly perform simulations for several values of the diameter and study the impact of dispersion on the first architectures based on CNTFET.

Figure 4 shows the values obtained for a coaxial grid with h=3nm and $\varepsilon_{ox}$=25. The values of these parameters alpha are the results of numerical approximations: they do not have any physical meaning and this explains their abrupt variations. At this stage, a polynomial approach to obtain relations $\alpha_0(d)$, $\alpha_1(d)$, and $\alpha_2(d)$, can't be done with accuracy. Thus either the main program of CNTFET-based circuit simulation calculates itself the alpha values or they are provided by tables in a circuit simulator.

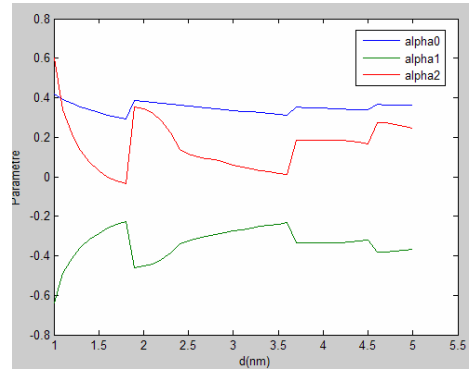

**Figure 4.** $\alpha_0$, $\alpha_1$ and $\alpha_2$ in function of d for a coaxial gate with h= 3 nm and $\varepsilon_{ox}$ =25 ($\varepsilon_{ox}$ = 18 – 40 for ZrO2).

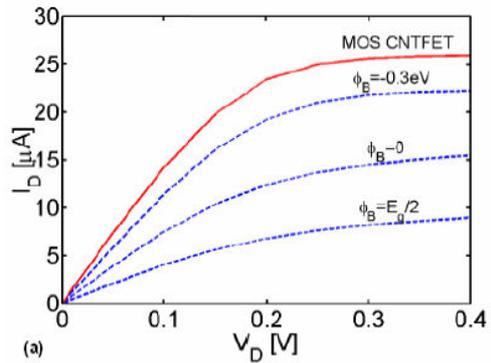

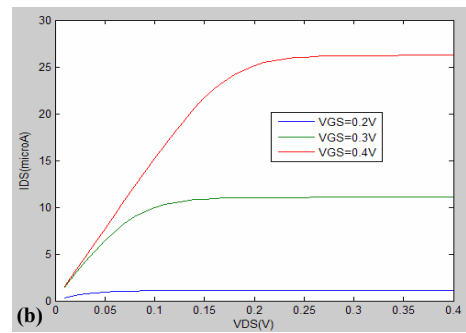

**Figure 5.** $I_{DS}$ in function of $V_{DS}$ for d=1nm, h=2nm and $\varepsilon_{ox}$=25 (a) Physical simulation results (b) Matlab CMVD simulations.

## 3. SIMULATION RESULTS

In this part, we present our first simulation results of the static characteristics for the ballistic transistor CNTFET n+-i-n+. These characteristics are obtained with our program in MATLAB. The aim is to compare them with the curves available in the literature. The publications of measurement results on CNTFET n+-i-n+ structures are nearly non-existent and the curves found in the literature are rather the results of numerical simulations from physical equations (1D Poisson equation and Schrodinger equation are self consistently solved).

### 3.1. Simulation $I_d(V_D)$ for $V_G$=0.4V with d=1nm

The curves shown in figure 5(a) are extracted from the article [1] whose aim is to show that MOS like CNTFET has better performances than SB CNTFET. A CMVD Matlab simulation was done with the same assumptions: coaxial gate with h=2nm, $\varepsilon_{ox}$=25 and d=1nm. Results are reported in figure 5(b) and we note a good accordance between the two curves concerning the MOS like CNTFET with $V_{GS}$=0.4V above conduction threshold.

### 3.2. Simulation for d=1.4nm, h=3nm and $\varepsilon_{ox}$ =25

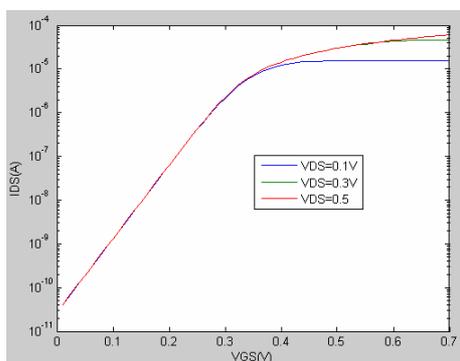

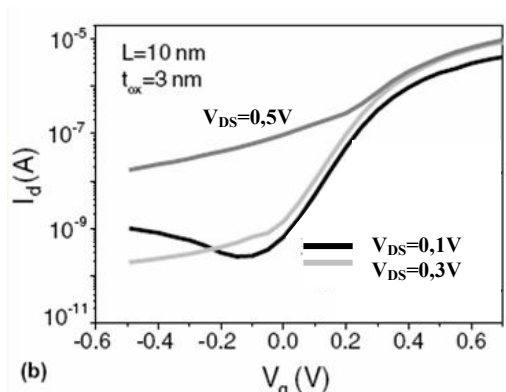

**Figure 6.** $I_{DS}$ in function of $V_{GS}$ for d=1.4nm, h=3nm and $\varepsilon_{ox}$=25 (a) Matlab CMVD simulations (b) Physical simulation results.

The curves of the figure 6(b) are extracted from [2] and they are the results of numerical simulations from physical equations. We compare to CMVD Matlab simulation, figure 6(a), obtained in the same bias conditions and with the same assumptions on geometry. For $V_{DS}$ equal to 0.1V or 0.3V and $V_{GS}$>0, the curves are comparable. But our model seems to give better performances:

| Parameter\Model | CMVD | Physical |
|---|---|---|
| Current $I_{on}$ | several 10μA | 10 μA |
| Inverse sub-threshold slope S | 60 mV/dec | 100 mV/dec |

**Tableau 1. Performances obtained with two types of modelling.**

The performances obtained with the compact model correspond to a highest limit of the totally ballistic CNTFET. In particular, the compact model does not take into account the length of the nanotube. For $V_{DS}$=0.5V, the curves don't match because there is another phenomenon not taken into account in our model. There is a flux of holes from the drain due to band to band tunneling (BTBT) and the off state is degraded.

## 4. DISPERSION IMPACT

### 4.1. Dispersion due to the diameter variations

When the CNTFETs are manufactured, there is a great dispersion on the obtained diameters. For example, for a required diameter of 1.4 nm, we obtain a collection of nanotubes between 1.2 and 1.6 nm. The object of this part is to characterize resulting dispersion on the main characteristics of the CNTFET. The currents at the one and off state correspond respectively to $V_{GS}$=0.7V and $V_{GS}$=0.1V, the inverse sub-threshold slope S is the minimal inverse slope observed.

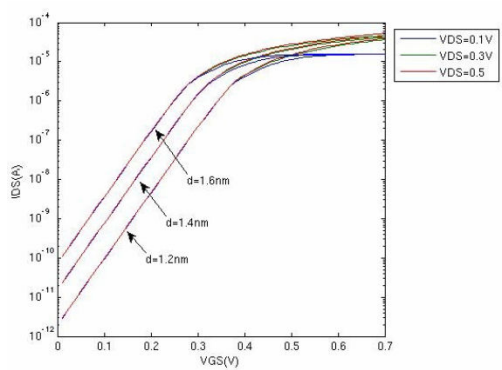

**Figure 7.** $I_{DS}$ in function of $V_{GS}$ for d=1.2nm, 1.4nm and 1.6nm, h=3nm and $\varepsilon_{ox}$=25.

According to the results shown on figure 7, we note that the sub-threshold slope is not modified by a diameter spreading. On the other hand, the current levels change: numerical values are reported on table 2. It shows that there is a limited dispersion on the $I_{on}$ current of approximately 10%; on the other hand, the $I_{on}/I_{off}$ ratio is very sensitive to diameter spreading with a factor 30 between the minimal value and the maximum value.

| Parameter\diam. | 1.2nm | 1.4nm | 1.6nm |
|---|---|---|---|
| Current $I_{on}$ | 37µA | 42µA | 45µA |
| Ratio $I_{on}/I_{off}$ | $379.10^3$ | $55.10^3$ | $12.10^3$ |

**Tableau 2. Simulated performances for several diameter values for $V_{DS}$=0.3V.**

A compromise should be found between the different required performances: a small diameter leads to a low off-state current and a high $I_{on}/I_{off}$ ratio while a large diameter leads to a high on-state current. And it is however difficult to control the diameter of the nanotube at the time of its manufacture in order to obtain the optimal value of 1.4nm.

Others physical parameters values vary in a wide range when nanotubes are manufactured. In the two following parts, we propose to study the impact of the variations of the Carbon-Carbon (C-C) bonding energy $V_\Pi$ and of the oxide dielectric constant $\varepsilon_{ox}$. The studied transistor has the following characteristics:

- nanotube diameter of 1.4nm corresponding to the chirality (15;5)
- coaxial geometry: the nanotube is surrounded by the gate
- gate oxide thickness of 3 nm
- oxide dielectric constant of 25 (plus or minus 10)
- C-C bonding energy of 3eV (plus or minus 0.3eV)

### 4.2. Dispersion due to the C-C bonding energy

C-C bonding energy depends, amongst others things, of nanotube geometry, of temperature and also of the possible topological defects. For example, the network of hexagons can contain pentagon-heptagon pairs leading to changes of the nanotube diameter and of the electronic behaviour around the Fermi level.

So we studied the impact of $V_\Pi$ variations around 3eV. We plotted $I_{DS}$-$V_{GS}$ and $I_{DS}$-$V_{DS}$ curves for several values of $V_\Pi$ and we determinated resulting variations on static performances of CNTFET.

As seen in the figures 8, 9 and 10, the static performances are modified: the on current decreases by 1.1 and the off current decreases by a factor 10 when $V_\Pi$ increases from 2.7 to 3.3 eV, which induces threshold voltage and $I_{ON}/I_{OFF}$ ratio increases. This last is multiplied by 8 when $V_\Pi$ moves from 2.7eV to 3.3 eV.

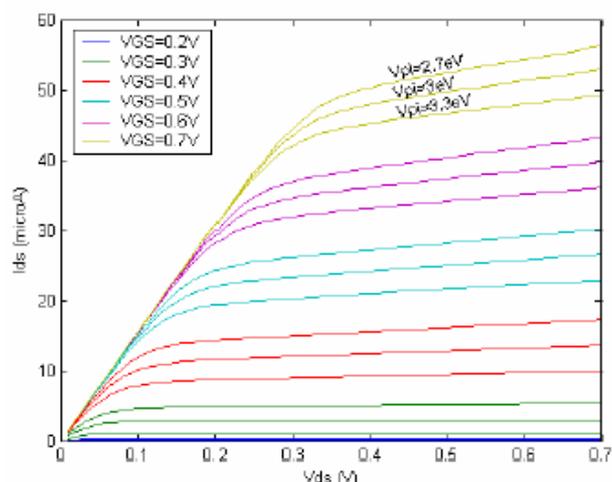

**Figure 8. Curves $I_{DS}$-$V_{DS}$ for $V_\Pi$ =2.7; 3.0; 3.3 eV.**

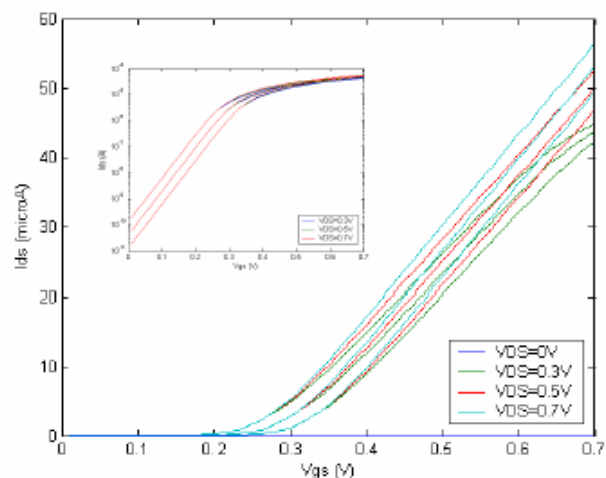

**Figure 9. Curves $I_{DS}$-$V_{GS}$ for $V_\Pi$ =2.7; 3.0; 3.3 eV with linear scale and logarithmic scale in the inset.**

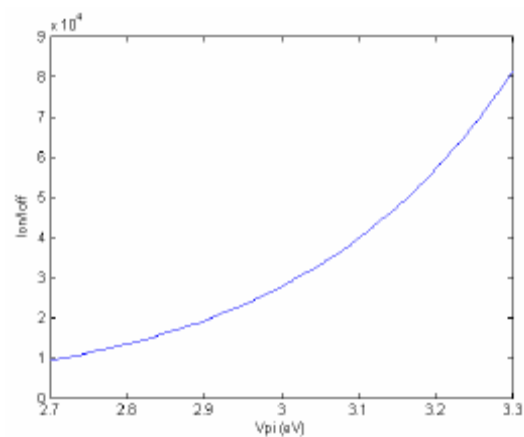

**Figure 10. Dispersion of $I_{ON}/I_{OFF}$ ratio in function of $V_\Pi$ =2.7.**

### 4.3. Dispersion due to the oxide dielectric constant

An oxide often used in the transistors for its great relative permittivity is the zirconium oxide also called zircon ($ZrO_2$). Its permittivity can vary to a significant degree, between 15 and 35. By studying dispersion using a Matlab routine, one obtains the following curves:

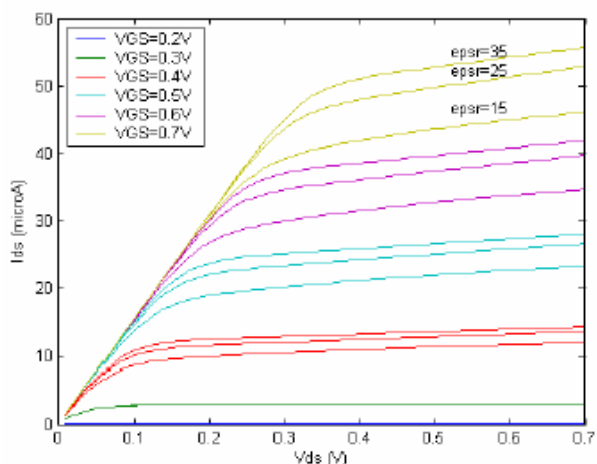

**Figure 11. Curves $I_{DS}$-$V_{DS}$ for $\varepsilon_r$ =15; 25; 35.**

On these curves giving $I_{DS}$ in function of $V_{DS}$, one realizes that the on current increases with $\varepsilon_r$. At $V_{DS} = 0,5$ V and $V_{GS} = 0,7$ V, for $\varepsilon_r$ varying from 15 to 35, the current $I_{ON}$ moves from 44 µA to 53 µA, that represents an increase of approximately 20 %.

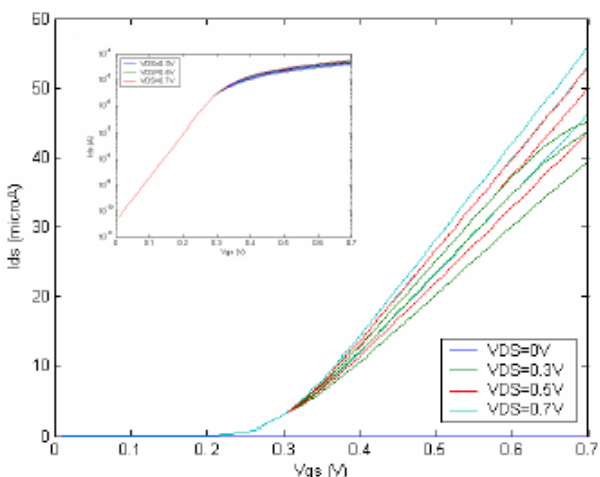

**Figure 12. Curves $I_{DS}$-$V_{GS}$ for $\varepsilon_r$ =15; 25; 35 with linear scale and logarithmic scale in the inset.**

The curves $I_{DS}$ in function of $V_{GS}$ in logarithmic scale show that the off current and the inverse sub-threshold slope are not modified. The linear scale show the current fluctuations at the on state previously noted, and the voltage threshold does not vary.

### 5. CONCLUSIONS

The compact modelling of the ballistic CNTFET whose source and drain are in doped nanotube is interesting because it corresponds to the most powerful transistor with nanotube. The model, based on a surface potential integration, is already exploitable provided that its validity range is well defined. In particular, the model is valid for low applied voltages $V_G$ in order to not populate the sub-bands of energy higher than 0.5eV. We added in our model implementation a module of automatic calculation for the parameters $\alpha_0$, $\alpha_1$ and $\alpha_2$ depending on the nanotube diameter in order to study the influence of dispersions on the CNTFET performances. The continuation of our work will both address the model improvement: extension of the validity field for the static model and addition of the dynamic behaviour [5] and the study of the dispersion influence on analogical and mixed architectures.

### 6. ACKNOWLEDGEMENT

The authors thank the partners of project NANOSYS: http://nanosys.ief.u-psud.fr/

This project is financed by the CNRS within the framework of an Inciting Joint Action Nanosciences.